\documentclass[10pt,showpacs,twocolumn]{revtex4}
\usepackage{graphicx}
\usepackage{amsmath}
\usepackage{amssymb}
\usepackage{}

\begin{document}
\title{Reply on the comment on ``Classical Simulations Including Electron Correlations for Sequential
Double Ionization''}
\author{Yueming Zhou, Cheng Huang, Qing Liao and Peixiang Lu}

\maketitle

In our Letter \cite{Zhou}, we studied sequential double ionization
(SDI) of Ar by the elliptically polarized laser pulses. In Ref.
\cite{Pfeiffer}, Pfeiffer {\it et al.} shown that the
independent-tunneling theory failed in predicting the ionization
time of the second electron. Meanwhile, the quantum calculation of
the fully correlated two-electron atom at the experimental
conditions is currently not feasible. Thus, we resorted to a
classical method. With a classical correlated model, we
demonstrated that the experimentally measured ionization times for
both electrons are quantitatively reproduced. Because of the
autoionization of the classical two-electron system, in our Letter
\cite{Zhou} we employed the Heisenberg-core potential to avoid
this problem. The Heisenberg-core potential is written as
\cite{Kirschbaum}:
\begin{equation}
V_H(r_i,p_i)=\frac{\xi ^2}{4\alpha r_i^2}exp\{\alpha[1-(\frac{r_i
p_i}{\xi})^4]\}
\end{equation}
where $r_i$ and $p_i$ are the position and the momentum of the $i$th
electron. There are two parameters in the Heisenberg-core potential:
$\alpha$ and $\xi$. In our letter, we chosen $\alpha=2$ and we
claimed that our results did not depend on $\alpha$.

In the preceding Comment \cite{Chandre}, Chandre {\it et al.} argued
that the excellent agreement between our calculation and
experimental data is a coincidental result of the parameters we
chose. Here, we demonstrate that our calculations are structurally
stable and do not depend on the parameter $\alpha$, and explain why
in the calculations of Chandre {\it et al.} \cite{Chandre} the time
delay between the two successive ionizations changes with $\alpha$.
As addressed in our Letter \cite{Zhou}, a necessary condition for
quantitative description of SDI is that the first and the second
ionization potentials of the model atom should be matched with the
realistic target. In our classical model, this condition is
satisfied by the combined action of the parameters $\alpha$ and
$\xi$. In fact, the parameter $\xi$ is not free when $\alpha$ is
given. It is set to make the minimum of the one-electron Hamiltonian
$H=\frac{-2}{{\bf r}_1}+\frac{{\bf p}_1^2}{2}+V_H{(r_1,p_1)}$ equal
to the second ionization potential of $Ar^+$. It can be determined
by eq. (3) of \cite{Chandre}. For Ar, the value of $\xi$ is 1.259
when $\alpha=2$. In our Letter \cite{Zhou}, we chosen $\xi=1.225$
because for $\xi=1.259$ one could not place the two electrons in
phase space with the ground-state energy of Ar (-1.59 a.u). For the
parameter $\xi=1.225$ and $\alpha=2$, the second ionization
potential of the model atom is -1.065 a.u., very close to second
ionization potential of $Ar^+$ (-1.02 a.u.). This small difference
of ionization potential between our model and the realistic target
has negligible influence on our results. In order to keep the second
ionization potential unchanged for different values of $\alpha$, the
value of $\xi$ should be adjusted. Figure 1(a) shows the value of
$\xi$ that keeps the second ionization potential (-1.065 a.u.)
unchanged for different values of $\alpha$. Figure 1(b) shows the
time difference between the ionizations of the two electrons for
different values of $\alpha$ [where the corresponding value of $\xi$
is shown in Fig. 1(a)]. Obviously, the results do not change with
the parameter $\alpha$.

\begin{figure}
\begin{center}
\includegraphics[width=9.5cm,clip]{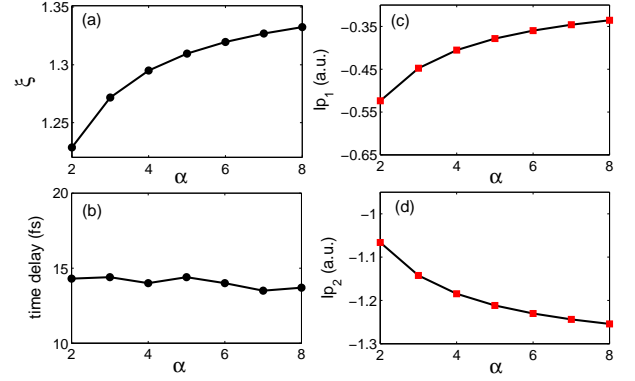}
\caption{\label{FIG. 1} (a) The value of $\xi$ that keeps the
ionization potentials unchanged when $\alpha$ varies. (b) The time
delay between the two successive ionizations in SDI for the
parameters shown in (a). The laser intensity is 4.0 PW/cm$^2$ and
the pulse duration is 33 fs. (c)(d) The first and second ionization
potentials as a function of $\alpha$ for the keeping $\xi=1.225$.}
\end{center}
\end{figure}

In \cite{Chandre}, Chandre {\it et al.} changed $\alpha$ while kept
$\xi$ unchanged. In this treatment, the first and second ionization
potentials change significantly as $\alpha$ varies, as shown in
figs. 1(c) and 1(d). When $\alpha=2$, the ionization potentials
nearly equal to the realistic target. However, when $\alpha$ becomes
larger, the first ionization potential increases and the second
ionization potential decreases. Naturally, the time difference
between the ionizations of the two electrons will increase as
$\alpha$ increases.

Thus, the change of the time difference for different values of
$\alpha$ in \cite{Chandre} originates from the shifting of the first
and second ionization potentials of the model atom. In order to
describe SDI accurately, it is necessary to make the ionization
potentials of the model atom match with the investigated target in
any case. In our model \cite{Zhou}, the parameters $\alpha$ and
$\xi$ should appear in pair to make the ionization potentials
unchanged for various value of $\alpha$. The results are
structurally stable when this condition is satisfied.

As addressed in our Letter \cite{Zhou}, the Heisenberg-core
potential in our model was added to (I) make the first and the
second ionization potentials match with realistic target and (II)
avoid autoionization while keep the two electrons being fully
correlated during the entire ionization process, which enables us to
investigate the multielectron effect in strong field ionization (see
Ref. \cite{Zhou2} for example). Recently, the ionization times of
both electrons in SDI are also reproduced by the soft-core potential
model when the ionization potentials are artificially adjusted to
those of the target \cite{Wang}. It indicates that the success of
the classical methods do not depend on the details of potential,
making it easy to be accepted that our calculations are stable upon
the parameters of the Heisenberg-core potential.

The detail of this issue was detailedly addressed in our recent
paper \cite{zhou3}.

\parskip=1\baselineskip
Yueming Zhou$^1$, Cheng Huang$^1$, Qing Liao$^1$ and Peixiang
Lu$^{1,2}$ \\$^1$Wuhan National Laboratory for Optoelectronics and
School of Physics, Huazhong University of Science and Technology,
Wuhan 430074, P. R. China \\$^2$Key Laboratory of Fundamental
Physical Quantities Measurement of Ministry of Education, Wuhan
430074, P. R. China


\begin{thebibliography}{99}

\bibitem{Zhou}
Y. Zhou, C. Huang, Q. Liao, and P. Lu, \prl {\bf 109,} 053004
(2012).

\bibitem{Pfeiffer}
A. N. Pfeiffer, C. Cirelli, M. Smolarski, R. D\"{o}rner, and U.
Keller, Nature Phys. {\bf 7,} 428 (2011).

\bibitem{Kirschbaum}
C. L. Kirschbaum and L. Wilets, Phys. Rev. A {\bf 21,} 834 (1980).

\bibitem{Chandre}
C. Chandre, A. Kamor, F. Mauger, and T. Uzer, comment on ``classical
simulations including electron correlations for sequential double
ionization".

\bibitem{Zhou2}
Y, Zhou, C. Huang, and P. Lu, Opt. Express {\bf 20,} 20201 (2012).

\bibitem{Wang}
X. Wang, J. Tian, A.N. Pfeiffer, C. Cirelli, U. Keller and J.H.
Eberly, arXiv:1208.1516v1 (2012).

\bibitem{zhou3}
Y. Zhou, Q. Zhang, C. Huang, and P. Lu, Phys. Rev. A {\bf 86,}
043427 (2012).

\end{thebibliography}
\end{document}